\documentclass[twocolumn,showpacs,amsmath,amssymb]{revtex4}
\usepackage{graphicx}
\usepackage{slashed}
\usepackage{pifont}
\usepackage{color}

\newcommand{\E}[1]{\times 10^{#1}}
\newcommand{\beqn}{\begin{equation}}
\newcommand{\eeqn}{\end{equation}}
\newcommand{\eps}{\epsilon}

\newcommand{\Real}{\mathtt{Re}}
\def\lambe{\ensuremath{{\mathchar'26\mkern-9mu \lambda_e}}}
\def\epsp{\ensuremath{\eps_{\perp}}}
\def\dumdt{\ensuremath{\frac{d\langle u_m \rangle}{dt}}}

\def\epslr{\ensuremath{\eps_{l,r}}}

\begin{document}
\title{Vacuum Decay Time in Strong External Fields}
\author{Lance Labun}
\author{Johann Rafelski}
\affiliation{Department of Physics, University of Arizona, Tucson, Arizona, 85721 USA}
\affiliation{Department f\"ur Physik der Ludwig-Maximillians-Universit\"at M\"unchen und\\
Maier-Leibniz-Laboratory, Am Coulombwall 1, 85748 Garching, Germany}
\date{22 October 2008} 

\begin{abstract}
We  consider dynamics of  vacuum decay and particle production in the context of
short pulse laser experiments.  We identify and evaluate the invariant
``materialization time,'' $\tau$,  the timescale for the conversion 
of an  electromagnetic field energy into particles, and we  compare to the laser related
time scales.
\end{abstract}


\maketitle

In the past decade high intensity short pulse  laser technology has advanced rapidly~\cite{Mourou}, 
pulses  achieved intensities of $10^{26}$ W/m$^2$~\cite{Hegelich, Hercules}.  
With subsequent concentration by coherent harmonic focusing allowing a further gain 
in intensity of around six orders of magnitude~\cite{Gordienko05}, laser technology is 
nearing the scale of rapid vacuum instability, $c\eps_0E_0^2/4\pi=4.65\E{33}$ W/m$^2$, 
where $E_0\equiv m^2c^3/e\hbar=1.32\E{18}\mathrm{V/m}$.
The study of vacuum instability with  laser pulses involves  dynamics
on a timescale set by the pulse length, which at optical 
frequencies implies that the fields are in existence for $\sim 10^{-15}$s, and 
may reach  $\sim 10^{-18}$s when coherent harmonic  focusing is used. 

The vacuum state of quantum electrodynamics (QED) is metastable in the presence of
electrical fields of any strength, but only in proximity  of $E_0$ does the effect occur
on an observable time scale~\cite{EH36,Schwinger51}, as we exhibit below.  
Specifically, we investigate whether the laser pulse 
timescale allows the vacuum in strong fields to relax, 
thereby admitting the new vacuum to experimental investigation
using pulsed lasers.  The dynamics of `false' vacuum decay have been studied in the context 
of spontaneous positron creation in heavy ion collisions~\cite{Pieper69,Mueller72,Rafelski74} and
cosmological models~\cite{Coleman77, Linde, Kuzmin}.  The QED vacuum decay has
not been directly observed in heavy ion collision experiments, due to the relatively long time 
scale of vacuum decay dynamics as compared to competing processes. However, particle
production in strong fields has found a fertile field in 
quantum chromodynamics~\cite{Casher79,Anderson}.

Considerable effort went into generalizing the Euler-Heisenberg-Schwinger (EHS)~\cite{EH36,Schwinger51} 
pair production mechanism for a variety of large-scale 
(compared to $\lambe=\hbar /m_ec=3.86\E{-13}\mathrm{m}$)
space- and time-dependent field  configurations~\cite{Brezin70,Narozhnyi70, Kim06}
and to incorporating back reaction~\cite{Mueller75, Casher79, Cooper93, Kluger98, Tomaras2000}.
A stable, modified vacuum state has only been obtained when the field fills a finite 
space-time domain~\cite{Mueller75}.  The perturbative vacuum is also stable for an ideal plane wave 
(laser) field of arbitrary strength, and thus many investigations have focused on
understanding pair production in optimized pulsed laser field 
configurations~\cite{Bunkin69,l1,l2,l3,l4,l5,l6,l7,Narozhnyi04,l8,l9}.
More recently it has been also noted that in the interaction of laser pulses with 
thin foils, the charge seperation effect due to a much greater electron mobility 
helps in  achieving  longitudinal electrical fields  of comparable strength as are present in
the laser pulse, a phenomenon  used in laser-ion  acceleration~\cite{Klimo08}.
We thus address in this work the general circumstance of a spatially homogeneous electrical field. 

In all laboratory experiments supercritical fields (fields capable of spontaneous particle 
production) will be strongly time dependent. One must distinguish two 
profoundly different experimental regimes involving vacuum rearragement: \\
a) the field is established on a time scale much faster 
than the typical vacuum decay time, such as was 
studied in heavy ion collisions~\cite{Mueller72,GMR85},  with
the spectrum of produced  particles (positrons) representative of the  single particle 
states achieved with the ultimate field strength;\\
b) the field is established on a time scale slower than the
 decay time of the vacuum (adiabatic switching). In this case a
particle  will be produced just upon achievement of supercriticality, and thus 
at zero (longitudinal) momentum. This is the assumption under which 
the  EHS instability is derived. \\
In both cases the  Fourier-frequency spectrum of the field formation assists the 
process of vacuum emission of particles, in which case we speak of
induced (as compared to spontaneous) vacuum 
decay~\cite{Soff77, Reinhardt81, Heben08, Schutzhold08}. 

We evaluate the rate per unit of time and volume of field energy materialization 
in the adiabatic EHS switch-on limit by calculating 
the tunneling probability in the presence of the local
potential generated by a (nearly) constant field~\cite{Casher79}.
One starts with the action of an electron with transverse energy $\eps^2_{\perp}=p_{\perp}^2+m^2$
in the inverted (i.e. Euclidean time) potential
\beqn \label{action} S = \int_0^{{\eps_{\perp\!}}/{eE}} \!\!\!\!\vert \vec p \vert dz 
            = \int_0^{{\eps_{\perp}\!}/{eE}} \!\!\!\!\!dz \sqrt{\eps_{\perp}^2-(eEz)^2} 
            = \frac{\pi\eps^2_{\perp}}{4eE}, \eeqn
with the upper bound determined by the turning point in the potential, at which the 
quasi- longitudinal momentum (and therefore the pair) becomes real.  The tunneling 
probability for the electron-positron pair is then twice the action in Eq.\,(\ref{action})
\beqn\label{Gamma} 
\Gamma(\eps_{\perp}^2) = \vert \exp(-2S) \vert^2 = e^{-\frac{\pi}{eE}\eps_{\perp}^2}.
 \eeqn
By integrating  $\prod_s \prod_{\epsp} [1-\Gamma(\eps_{\perp}^2)]$ over all $s,p_\perp$ 
one reproduces the Schwinger  series expression for  the total vacuum persistence probability, 
as was first noted by Nikishov~\cite{Nikishov69}.  However, to obtain the rate 
$d\langle N \rangle/dtdV$, we sum the probability Eq.~\eqref{Gamma} over the volume-normalized 
density of states for a spin-1/2 particle $2_sd^3p/(2\pi)^3$. 
With the assumption of adiabaticity, the longitudinal momentum is determined entirely by the 
electric field~\cite{Cohen08}, so that $dp_3/dt=eE$, and 
\beqn \label{dNdt}
 W=\frac{d\langle N \rangle}{dtdV} = 2_s\int \!\frac{d^3p}{(2\pi)^3dt}~\Gamma(\epsp^2) 
      = \frac{\alpha}{\pi^2}E^2e^{-\frac{\pi E_0}{E}}.
 \eeqn  

We obtain the rate at which energy is transferred to EHS pairs by 
including in Eq.\,\eqref{dNdt} the transverse energy of the electron 
and positron before integrating over the momentum
\beqn\label{rate-integ} 
\frac{d\langle u_m\rangle }{dt} =
 2_s\int \frac{d^2p_{\perp}}{(2\pi)^2}\frac{dp_3}{2\pi dt} (\eps_++\eps_-) \Gamma(\epsp^2),
\eeqn
in which $\eps_+~(\eps_-)$ is the asymptotic energy of the electron (positron).  As we 
explicitly assume 
that only one pair is created per event, momentum conservation allows us to put
 $\eps_+ =\eps_- = \sqrt{p^2_{\perp}+m^2}$.  Thus, the rate of energy conversion becomes
\beqn 
\frac{d\langle u_m \rangle}{dt} = \frac{eE}{2\pi^2}\int_m^{\infty} d\eps_{\perp} 
	\,2\eps^2_{\perp} e^{-\frac{\pi}{eE}\eps_{\perp}^2}\ .
\eeqn
Integration by parts results in
\beqn \label{Erate}
 \frac{d\langle u_m \rangle}{dt} = \omega_0 E^2e^{\frac{-\pi E_0}{E}}
   \left\{ 1+ h\!\left(\!\sqrt{\frac{\pi E_0}{E}}\right) \right\} ,
\eeqn 
in which $\omega_0 := \alpha c/\pi^2\lambe=\alpha mc^2/\pi^2\hbar=5.740\E{17}$s$^{-1}$
and $h(z) :=  {\sqrt{\pi}}e^{z^2}\mathrm{erfc}(z)/{2z}.$ 
The asymptotic behavior of the complementary error function implies
 that $h(z)$ increases linearly with $E/E_0$, specifically, for $z \gg 1$, $h(z)\sim E/4E_0$. 

The relaxation time of the metastable with-field vacuum state via materialization
is the ratio of the available supply of field energy $u_f$ (density) to 
the rate of electromagnetic field energy conversion Eq.\,(\ref{Erate}):
\beqn\label{taudef}
 \tau := u_f\left(\dumdt\right)^{-1}.
 \eeqn
We assume that the pairs decohere rapidly so that the reverse process is impossible, 
and Eq.\,\ref{taudef} then generates the usual kinetic result
\beqn 
u(t) = u(t_0)e^{-\int_{t_0} ^t\frac{dt'}{\tau}}. \nonumber 
\eeqn
 $\tau$ provides the time at which all field energy is converted into mass, 
and hence we refer to it as the \emph{materialization time} of the field.

\begin{figure}
\includegraphics[scale=.35]{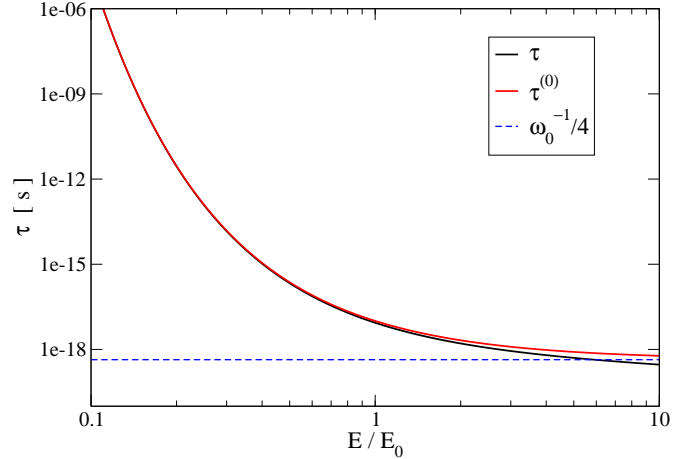}
\caption{Materialization time $\tau$ (solid line)  and  $\tau^{(0)}$  (lighter, red) as a 
function of the external normalized field $E/E_0$. Dashed line:  
$\omega_0^{-1}/4 = \pi^2\hbar/4\alpha mc^2 =0.435$\,as .\label{pureEtau}}
\end{figure}

A rough time scale may be obtained by ignoring the second term Eq.~\eqref{Erate} and the 
nonlinear corrections in the field energy, using $u_f = E^2/2$:
\beqn\label{tau0} 
\tau^{(0)} := \frac{\omega_0^{-1}}{4}e^{\frac{\pi E_0}{E}} .
\eeqn
This is the central result of this paper in its simplest qualitative form, 
shown in figure \ref{pureEtau} as the upper (red) line.
Materialization time for $E<E_0$ is longer than 8 as. 

The lifetime of the field is directly related to the quantity of (usable) energy present, 
and positive or negative corrections to the energy density translate into increases or decreases
in $\tau$.  The nonlinear contributions are governed by the real part of the effective Lagrangian via 
\beqn \label{udef} 
u_f = E\cdot D - \Real \mathcal{L},\quad\textrm{where}~D \equiv \frac{\partial }{\partial E}\Real \mathcal{L}.\nonumber
 \eeqn 
Using the resummed expression of~\cite{Mueller77}, $u_f$ may be calculated explicitly.  
Reference \cite{Valluri} displays contour plots for parallel and perpendicular electromagnetic 
fields up to $5E_0$ using expansions of the integrated expression~\cite{Cho00}, showing that 
the correction $\vert \Delta u_f \vert$ never exceeds $10^{-2}E_0^2$.  More importantly, 
the correction is positive for pair producing field configurations 
(for which no frame exists such that $E\equiv 0$), 
implying that the field should persist longer than expected based on the linear Maxwellian 
formula.  However, the effect is over-compensated by the correction in Eq.\,(\ref{Erate}) 
introduced by proper weighting of the phase space integration.
In figure \ref{pureEtau}, the resulting materialization time $\tau$ is shown (solid lower 
black line) for a constant, purely electric field. 

We next evaluate $\tau$ adding a constant, homogeneous 
magnetic field.  The energy available in presence of both electric and magnetic 
fields is evaluated in the local rest frame  
from the four-vector of energy-momentum $T^{\mu\nu}v_{\nu}$ (see e.g. \cite{Greiner92}) i.e.
$u_f=\vert T^{\mu\nu}v_{\nu}\vert$. This ``mass density'' of the field 
is expressed in terms of the invariants $\mathcal{F}=(1/4)F^{\mu\nu}F_{\mu\nu}=\frac{1}{2}(B^2-E^2)$ 
and $\mathcal{G}=(1/4)F^{\mu\nu}F^*_{\mu\nu}=E\cdot B$ as
\beqn \label{cov-u} u_f= 
\sqrt {(\mathcal{F}^2+\mathcal{G}^2)f^2(\mathcal{F},\mathcal{G})+A^2(\mathcal{F},\mathcal{G})},
 \eeqn  
where $f=\partial \mathcal{L}/\partial \mathcal{F} \to -1$ (Maxwell) and 
$A$ is the conformal anomaly induced by external fields, $T^\mu_\mu =4A$~\cite{Labun08}.

We remark at this point that $d\langle u_m\rangle / dt$ is Lorentz invariant.  It follows that $\tau$ as defined
is the Lorentz invariant (proper) decay time of the vacuum, and we 
may choose a suitable frame in which to evaluate $\tau$.  
When $E\cdot B\equiv 0$, pairs are produced whenever the generalized electric field 
$a=\sqrt{\sqrt{\mathcal{F}^2+\mathcal{G}^2}-\mathcal{F}}$ is nonzero, and the rate reduces to that of
Eq.~\eqref{tau0}.  For  $E\cdot B\ne  0$, a reference
frame exists where $E$ and $B$ are either parallel or antiparallel.  Carrying through 
the tunneling calculation in this frame, we observe the quantization of the transverse
momentum of the final states:  
\beqn \nonumber
\Gamma_{l,r} = e^{-\frac{\pi}{eE}\eps_{l,r}^2}\ ,
\eeqn
with $\eps_{l,r}^2$ the energy eigenvalues in the combined field,
\begin{equation}\label{Bevalues}
 \epslr^2 = \begin{cases} m^2+2eBl & r = -1, \\ m^2+2eB(l+1) & r=+1 ,
\end{cases}
 \end{equation}
in which $l=0,1,2,...$  The integration 
over transverse momenta converts into 
the usual sum over Landau levels,
\beqn \label{Landau-sum}
\dumdt 
    =  \frac{(eE)(eB)}{(2\pi)^2} \sum_{l,r} \eps_{l,r}e^{-\frac{\pi}{eE}\eps_{l,r}^2}\ .
 \eeqn
Noting the double degeneracy of all but the $l=0$ mode, we sum over $r$ and apply the Euler-Maclaurin 
summation formula.  
This produces the convergent form
\begin{eqnarray}\label{EB-dudt}
\dumdt\!\!\! &= &\!\!\!
\omega_0E^2e^{\!-\frac{\pi E_0}{E}}\!\left\{\!1\!+ h\!\left(\sqrt{\frac{\pi E_0}{E}}\right)\!- \chi(B^2,E)\!\right\}
\\ \chi& \!\!\!\!:=&\!\!\!\! \frac{E_0}{E} 
 \sum_{k=1}^{\infty}\frac{\mathfrak{B}_{2k}}{(2k)!}\!\left(\frac{2B}{E_0}\right)^{\!\!2k}\!\!\!
   e^{\frac{\pi E_0}{E}}\!\frac{d^{2k-1}}{dx^{2k-1}}\!\left[\!\sqrt{x}e^{-\frac{\pi E_0}{E}x}\!\right]_{x=1} 
   \nonumber
\end{eqnarray}
with $\omega_0$ and $h(z)$ defined as above and $\mathfrak{B}_{2k}$ the Bernoulli numbers.  
The ``x coth x'' found in the pair creation 
rate~\cite{Nikishov69, Kim06, Cohen08} has been replaced due to the weighting of the phase space 
integral, 
resulting in a dependence on the magnitude $B^2$.

\begin{figure}
\includegraphics[scale=.35]{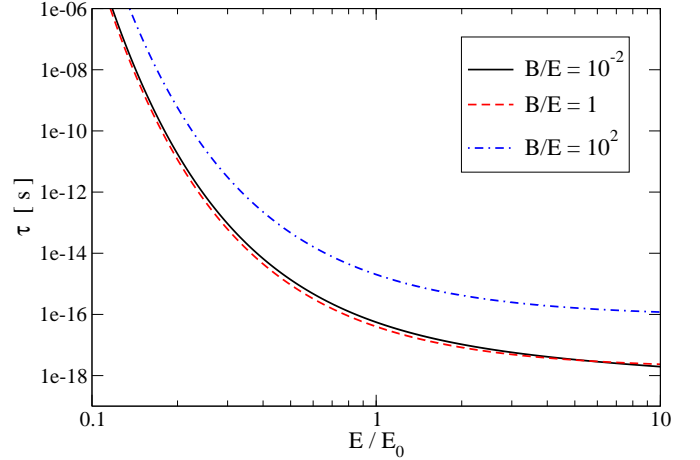}
\caption{
 $\tau$ for several ratios of $B / E$ as function of $E/E_0$, evaluated according to Eq.~\eqref{EBtime}, 
 using the full non-linear $u_f$ of Eq.~\eqref{cov-u}.    
 \label{EBtau}}
\end{figure}

Combining Eqs.\,\eqref{cov-u},\eqref{EB-dudt} the materialization time is
\beqn \label{EBtime} \tau = \frac{u_f}{E^2}\frac{\omega_0^{-1} e^{\left(\pi E_0/E\right)}} 
  {1+h\!\left(\!\sqrt{\frac{\pi E_0}{E}}\right)\! - \chi(B^2,E)}.
\eeqn
In figure~\ref{EBtau}, Eq.\,\eqref{EBtime} is evaluated for
$B/ E=10^{-2},1,10^2$.  For $B\gtrapprox E$ the lifetime of the field is increased over the 
pure electric case despite the augmented production rate evidenced by the coth factor 
mentioned above.

In table\,\,\ref{nums}, we exhibit a few points of 
reference for a pure electric field, listing the particle creation rate from Eq.\,\eqref{dNdt} 
and the expected lifetime of the field given complete materialization from Eq.\,\eqref{taudef}.
 At $E/E_0 = 0.2$ (4\% of critical power intensity) and a time scale of $10^{-15}$s for example, 
the materialization  
rate shows that $0.036\%$ of the field energy is converted and therefore approximately 280~nC of electrons
(positrons) created per $\mu\mathrm{m}^{-3}$, a signature which should be well observable.
If the field energy is in kJ range, this implies that 0.15 J materialize, which greatly
exceeds the energy converted into particles $O(10~\mathrm{erg})$ in most extreme laboratory particle 
collision reactions.  Thus, for experiments alert to pair production, actually attaining the 
critical field $E_0$ is not necessary for a clear signal of vacuum decay (materialization) processes.
In the context of relativistic focusing, where the characteristic time scale is expected to be as short
as a few attoseconds, the critical field strength needs to be reached to achieve spontaneous 
vacuum decay (as opposed to induced pair creation).

\begin{table}\caption{Materialization characteristics (yield rate $W$, 
relaxation time $\tau$) for specific applied fields. \label{nums}}
\begin{tabular}{c|c|c}
 $E/E_0$ & $W~[\mathrm{\mu m}^{-3}\mathrm{fs}^{-1}]$ & $\tau~[\mathrm{fs}]$\\ \hline
 0.0628 & 1 & $2.275\E{18}$  \\ 
0.1 & $3.102\E{8}$ & $1.88\E{10}$  \\
 0.2 & $8.234\E{15}$ & $2800$  \\
0.402 & $9.68\E{19}$ & $1$ \\
 1 & $5.903\E{22}$ & $8.85\E{-3}$
 \end{tabular}
\end{table}

We further note that in the above example
the percentage of the field energy converted will remain small, 
and materialization will not present a significant source of dissipation in practice.  However 
materialization of near-critical fields ($E/E_0~\sim~1$) obtained in 
relativistic focusing can lead to the formation of large $O(50~\mathrm{nm}^3)$ 
spatial domains of electron-positron-photon plasma with $T\simeq 2$ MeV, allowing 
experiments to test the strongly coupled regime of QED and provide
 an accessible analogy for the current interest in quark-gluon plasma~\cite{Inga08}.

In summary, we have studied the materialization  time $\tau$ of the 
 electromagnetic field in view of pair production at high field intensity/energy density.
We presented the field dependence of $\tau$ and found that a field of order
$0.2E_0$ is sufficient for observable materialization. Our current study relies
on an adiabatically changing field configuration, as is appropriate in the EHS context.  
This is consistent a posteriori for $E\to E_0$ given
the characteristic times for materialization  in critical fields, which are 1000 times
shorter than the typical intense pulse laser fields operating at 10 femtosecond scale.  
On the other hand, our results imply that stronger fields $\sim E_0$ and closer investigations 
are necessary for the much shorter (attosecond range) field pulses generated in relativistic focusing.
The here introduced concept of the field materialization
 time may be  generalized to such more intense and shorter lived field configurations, because  
these calculations can be undertaken within the same semiclassical approach. 

\acknowledgments 
This research was supported
by the U.S. Department of Energy grant DE-FG02-04ER4131
and by the DFG--LMUexcellent grant. We thank for his generous hospitality 
Prof. Dr. D. Habs, Director of the Cluster of Excellence in Laser Science ---
Munich Center for\ Advanced\ Photonics.

\end{document}